\documentclass[aps,a4paper,12pt]{revtex4}
\pdfoutput=1 
\usepackage{epsfig}
\usepackage{graphicx}
\usepackage{amsmath,amssymb,color}
\usepackage[english]{babel}

\parskip=\medskipamount

\newcommand{\eq}[1]{(\ref{#1})}
\newcommand{\fig}[1]{Fig.~\ref{#1}}

\newcommand{\be}{\begin{equation}}
\newcommand{\ee}{\end{equation}}

\newcommand\disp{\displaystyle}

\newcommand{\la}{\left<}
\newcommand{\ra}{\right>}
\newcommand{\eps}{\varepsilon}

\newcommand{\im}{\textrm{Im}\,}

\begin{document}

\title{Native ultrametricity of sparse random ensembles}

\author{V. Avetisov$^{1,2}$, P. L. Krapivsky$^{3}$ and S. Nechaev$^{4,5}$}

\address{$^1$N. N. Semenov Institute of Chemical Physics of the Russian Academy of Sciences,
119991, Moscow, Russia \\ $^2$Department of Applied Mathematics, National Research University
Higher School of Economics, 101000, Moscow, Russia \\
$^3$ Physics Department, Boston University, Boston MA 02215, USA \\
$^4$Universit\'e Paris-Sud/CNRS, LPTMS, UMR8626, B\^at. 100, 91405 Orsay, France \\
$^5$P. N. Lebedev Physical Institute of the Russian Academy of Sciences, 119991, Moscow, Russia}

\date{\today}

\begin{abstract}

We investigate the eigenvalue density in ensembles of large sparse Bernoulli random matrices. We
demonstrate that the fraction of linear subgraphs just below the percolation threshold is about
95\% of all finite subgraphs, and the distribution of linear chains is purely exponential. We
analyze in detail the spectral density of ensembles of linear subgraphs, discuss its ultrametric
nature and show that near the spectrum boundary, the tail of the spectral density exhibits a
Lifshitz singularity typical for Anderson localization. We also discuss an intriguing connection of
the spectral density to the Dedekind $\eta$-function. We conjecture that ultrametricity is inherit
to complex systems with extremal sparse statistics and argue that a number-theoretic ultrametricity
emerges in any rare-event statistics.

\end{abstract}

\maketitle

\section{Introduction}

\subsection{Some notions about ultrametricity}

The concept of ultrametricity is related to a special class of metrics. Generally, a metric space
is a set of elements equipped by pairwise distances. The metric $d(x,y)$ meets three requirements:
i) non-negativity, $d(x,y)>0$ for $x\neq y$, and $d(x,y)=0$ for $x=y$, ii) symmetry,
$d(x,y)=d(y,x)$ and iii) the triangle inequality, $d(x,z)\le d(x,y) + d(y,z)$. Ultrametric spaces
obey the strong triangular inequality, $d(x,z) \le \max\{d(x,y), d(y,z)\}$, allowing only acute
isosceles and equilateral triangles.

The strong triangle inequality radically changes the properties of spaces in comparison with our
intuitive expectations based of Euclidean geometry.  In ultrametric spaces, the sum of distances
does not exceed the largest summand, i.e. the Archimedean  principle does not hold. Few
counterintuitive properties are related to ultrametric balls: Every point inside a ball is its
center; the radius of a ball is equal to its diameter; there are no balls that overlap only
partially, if two balls have a common point, then one of them is completely embedded into the
other. The last property implies that any ultrametric ball can be divided into a set of smaller
balls, each of them can be divided into even smaller ones, etc. In other words, an ultrametric
space resembles not a line of segments, but a branching tree of hierarchically nested balls
(``basins"). The terminal points of a tree, i.e. the tree boundary, also obey the strong triangle
inequality: for any two such points, A and B, the distance to the root of the minimal subtree to
which these points belong, determines the distance between A and B. As a result, ultrametric geometry, in general, fixes taxonomic (i.e. hierarchical) tree-like
relationships between the elements.

Ultrametricity is closely related to the $p$-adic numerical norm, which originally arose in the
context of number theory \cite{KH}. The role of ultrametric spaces and $p$-adic numbers in number
theory \cite{Koblitz,Robert} and in other branches of mathematics, such as $p$-adic analysis
\cite{VVZ} and non-Archimedean analytic geometry \cite{VB}, is growing. More recently, there have
been numerous attempts to apply $p$-adic numbers and ultrametric spaces to various branches of
physics ranging from string theory to inflationary cosmology (see e.g. \cite{ST1,ST2,ST3,DKKV,Susskind,Van}).

Perhaps most naturally ultrametric spaces emerge in the realm of complex systems. Indeed, the
ultrametric description is inherently multi-scale, thus it is not surprising that ultrametricity
provides a fertile framework for describing complex systems characterized by a large number of
order parameters. For instance, signatures of ultrametricity were discovered in the context of spin
glasses (see \cite{SG:ultra} for an earlier review). Spin glasses are characterized by a huge
number of metastable states and the equilibration in spin glasses proceeds via tree-like splitting
of the phase space into hierarchically nested domains. When the temperature decreases, the most
distant groups of spin-glass states get factorized first. Then, each of these groups splits into a
number of subgroups, etc. The scales of hierarchically nested sets of equilibrated states satisfy
the strong triangle inequality, i.e. the multi-scale splitting of spin-glass phases obey
ultrametric relations.

The physical origin behind the emergence of ultrametricity in spin-glass systems is quite general.
In a complex system with an extremely large number of metastable states, the phase trajectories
cannot totally explore all states: they cover only a tiny part of phase space almost without
returns to already visited regions. For this reason, the factorization of spin-glass phases is
similar to random branching in a high-dimensional (infinitely-dimensional in the thermodynamic
limit) space. The factorization of spin-glass phases is specified only by branching points because
once two branches are dispersed in a high-dimensional space, they never join. The ultrametric
ansatz \cite{parisi,Parisi} proposed to describe the spin-glasses by infinite (in the thermodynamic
limit) number of order parameters, was proven in \cite{talagrand} to be an exact ground state of
the Sherrington-Kirkpatrick model \cite{SK75}.

Shortly after the formulation of the ultrametric ansatz in the context of spin glasses, similar
ideas were proposed for the description of native states of protein molecules \cite{nat_prot1,
nat_prot2}. Proteins are highly frustrated polymer systems characterized by rugged energy
landscapes of very high dimensionality. The ultrametricity of proteins implies a multi-scale
representation of the protein energy landscape by means of ranging the local minima into
hierarchically embedded basins of minima. Ultrametricity follows from the conjecture that the
equilibration time within the basin is significantly smaller than the time to escape the basin. As
a result, the transitions between local minima obey the strong triangle inequality. In such an
approximation, dynamics on a complex energy landscape can be modelled via a jump-like random
process, propagating in an ultrametric space of states. The approximation of the protein energy
landscape by a self-similarly branching tree of basins has been shown to be very fruitful for
describing a large body of experimentally justified features of protein conformational dynamics and
protein fluctuation mobility at temperatures covering very wide range from room temperatures up to
the deeply frozen states \cite{avet}.

With respect to ultrametricity and the self-similarity of protein energy landscapes, it is
important to keep in mind that proteins are functional systems. Proteins precisely operate at the
atomic level using sparse-event statistics, which is a bridge between non-local protein dynamics in
high-dimensional conformational space and low-dimensional movements of particular atomic units
related to protein functionality.

\subsection{Ultrametricity and rare-event statistics}

In the context of high dimensionality, randomness, and rare-event statistics, ultrametricity
emerges in diverse subjects, e.g., in genetic trees, in DNA folding (or crumpling) in chromosomes,
in the behavior of hierarchically organized ecological, social, and economic systems, etc. The fact
that ultrametricity is naturally rooted in high dimensionality, randomness, and sparse statistics,
has been unambiguously observed recently in \cite{zubarev}. It was proven that in a $D$-dimensional
Euclidean space the distances between points in a highly sparse sampling tends to the ultrametric
distances as $D\to\infty$.

In this work we bring another intriguing connection between rare-event statistics and
ultrametricity. We demonstrate that ultrametricity is inherit for ensembles of sparse topological
graphs. In particular, comparing a limiting behavior of spectral density of bi-diagonal random
matrices, with the behavior of the function $f(z)=\sqrt{-\log|\eta(z)|}$, where $\eta(z)$ is the
Dedekind $\eta$-function, and $z$ tends to rational points on real axis, we conjecture that the
eigenvalue distribution of sparse ensembles shares some number-theoretic properties related to
ultrametricity.

Our finding provides an example of a nontrivial manifestation of rare-event Bernoullian noise in
some collective observables in large systems (like spectral density, for example). From this point
of view, our work should warn researchers working with effects of ultra-low doses of chemicals: in
the statistical analysis of, say, absorption spectra of extremely diluted solutions of chain-like
polymer molecules, the valuable signal should be clearly purified from the background noise, which
itself could have a very peculiar ultrametric shape.

The paper is organized as follows. In the Section II we formulate the model under consideration and
describe the results of our numerical simulations. In what follows we pay attention to statistics
of linear subgraphs, presented by bi-diagonal matrices. Specifically, we derive the distribution of
the fraction of linear subgraphs at the percolation threshold in the Section III, and study the
spectral properties of Bernoulli ensembles of bi-diagonal matrices in the Section IV. The
number-theoretic properties of limiting spectral statistics in ensembles of random bi-diagonal
matrices are discussed in the Section V. The results of the work and the relation with other
natural systems possessing the number-theoretic properties under consideration, are summarized in
the Discussion.

\section{The model}

The Bernoulli matrix model is defined as follows. Take a large $N\times N$ symmetric matrix $A$.
The matrix elements, $a_{ij}=a_{ji}$, are assumed to be independent identically distributed binary
random variables such that non-diagonal elements are equal to one with probability $q$, while
diagonal elements vanish. Thus, $a_{ii}=0$, and for $i\ne j$ one has
\be
a_{ij}=
\begin{cases}
1 & {\rm with ~probability} ~~ q \\
0 & {\rm with ~probability} ~~ 1-q
\end{cases}
\label{eq:01}
\ee

The matrix $A$ can be regarded as an adjacency matrix of a random Erd\H{o}s-R\'enyi graph, ${\cal
G}$, without self-connections. The eigenvalues, $\lambda_n$ ($n=1,...,N$), of the symmetric  matrix
$A$ are real. Let $\rho(\lambda)$ be the eigenvalue density of the ensemble of such matrices. For
$N\gg 1$ the limiting shape of $\rho(\lambda)$ is known in various cases. If $q=\mathcal{O}(1)$,
the function $\rho(\lambda)$ tends to the Wigner semicircle law, $\sqrt{4N-\lambda^2}$, typical for
the Gaussian matrix ensembles. For $q=c/N$ ($c>1$), the matrix $A$ is \emph{sparse} and the density
$\rho(\lambda)$ in the ensemble of sparse matrices has singularities at finite values of $c$
\cite{rod1,rod2,fyod}. In Refs.~\cite{evan,bauer,sem,kuch}, the behavior of the spectral density,
$\rho(\lambda)$, has been analyzed near $c=1$ ($N\gg 1$). It has been pointed out that the function
$\rho(\lambda)$ becomes more and more singular as $c$ approaches 1 from above.

For ensemble of matrices $A$ the percolation transition occurs at the value $q_c=1/N$. (The true
transition occurs, certainly, in the thermodynamic limit, $N\to\infty$.) For $q>q_c$, the graph
${\cal G}$ has a giant component whose size is proportional to $N$, and numerous small components
(almost all of them are trees); for $q<q_c$, the graph ${\cal G}$ is the collection of small
components (almost all of them are again trees). For $q\ll q_c$ the isolated vertices dominate,
contributing to the trivial spectral density, $\rho(\lambda)=\delta(\lambda)$. The most interesting
region is close to $q_c$. As long as  $|q-q_c|=O(N^{-4/3})$, the giant component is still absent.
In what follows we discuss the behavior of the spectral density $\rho(\lambda)$ at $q$ in this
region; more precisely, we usually chose $q$ slightly below $q_c$.

\begin{figure}[ht]
\centerline{\includegraphics[width=14cm]{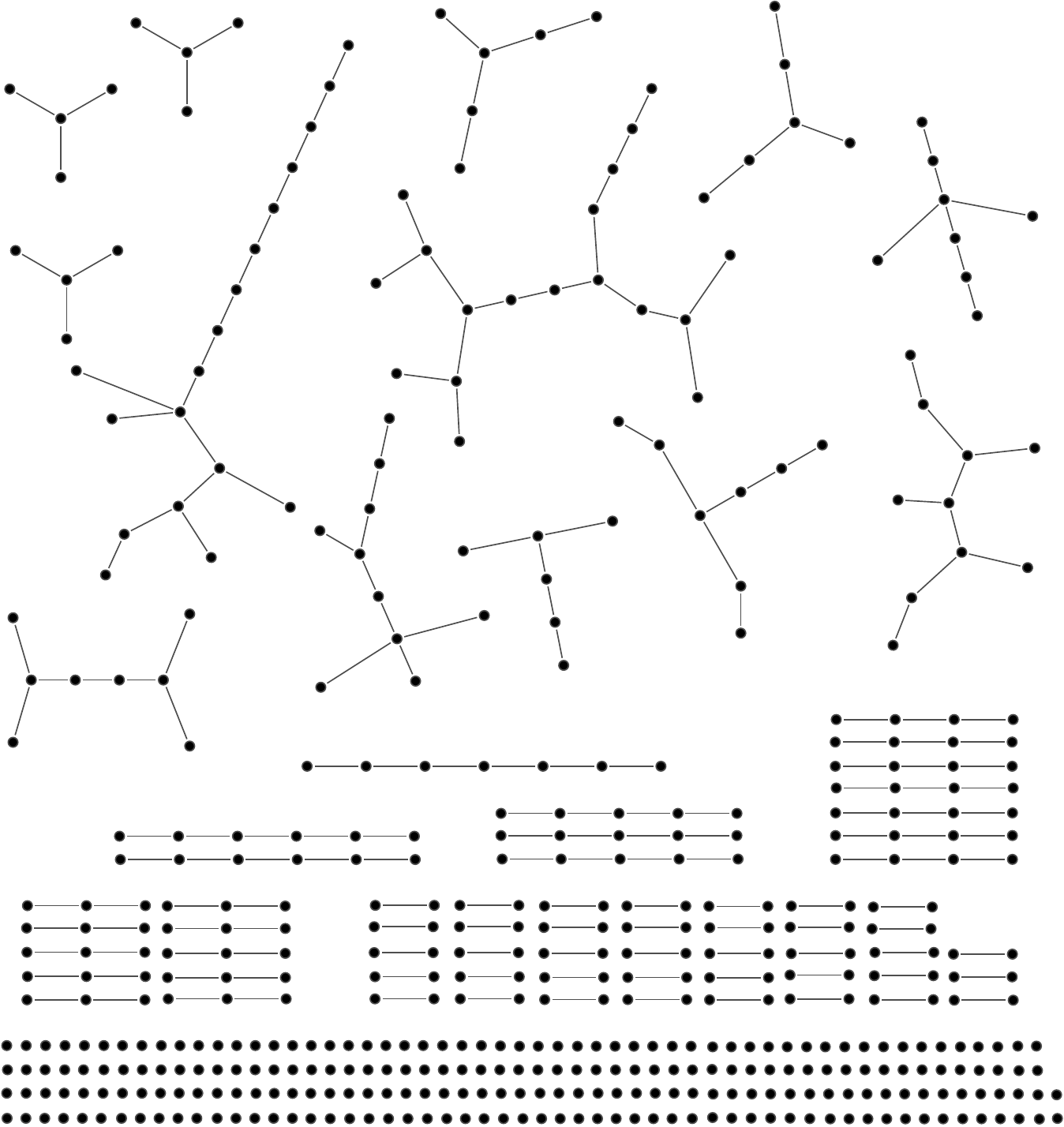}}
\caption{Typical sample of components constituting a random graph associated with an adjacency
matrix $A$ generated by the rule \eq{eq:01}; the parameters are $N=500$ and $q=2.0028\times
10^{-3}$.}
\label{fig:01}
\end{figure}

An example of a random graph with a randomly generated $N\times N = 500\times 500$ adjacency matrix
at $q=2.0028\times 10^{-3}$ is shown in \fig{fig:01}. Note that the components are predominantly
linear subgraphs. In other realizations we observed components with a single loop, but they are
rare, usually a few per realization; more complicated components, namely those with more than one
loop, appear very rarely.

The whole spectrum of a symmetric matrix $A$ in a sparse regime below the percolation point
consists of singular peaks coming from the graph components (maximal disjoint subgraphs
constituting the random graph). For a subgraph of size $k$, the corresponding adjacency matrix
$A_k$ of size $k\times k$ and the spectrum is determined by the equation $\det(A_k - \lambda I_k) =
0$, where $I_k$ is the $k\times k$ identity matrix. There are $k$ eigenvalues; apart from $k$, the
eigenvalues depend on the topology of the subgraph. The heights of peaks are the eigenvalue
multiplicities, the eigenvalue $\lambda=0$ appears with the highest multiplicity. For instance, the
graph shown in \fig{fig:01} has 222 subgraphs of size one (isolated vertices) each of them
contributing to $\lambda=0$. There are also 10 linear chains of length three, 3 chains of size
five, and 1 chain of size seven contributing to $\lambda=0$; generally any chain of odd length has
eigenvalue 0. Trees which are not linear chains also can have eigenvalue 0. One can bipartite any
tree into two disjoint parts $X_1$ and $X_2$ with edges only between $X_1$ and $X_2$. If the sizes
$n_1=|X_1|$ and $n_2=|X_2|$ of the parts are different, say $n_1>n_2$, the adjacency matrix has
eigenvalue 0 with multiplicity at least $n_1-n_2$.

\begin{figure}[ht]
\centerline{\includegraphics[width=15cm]{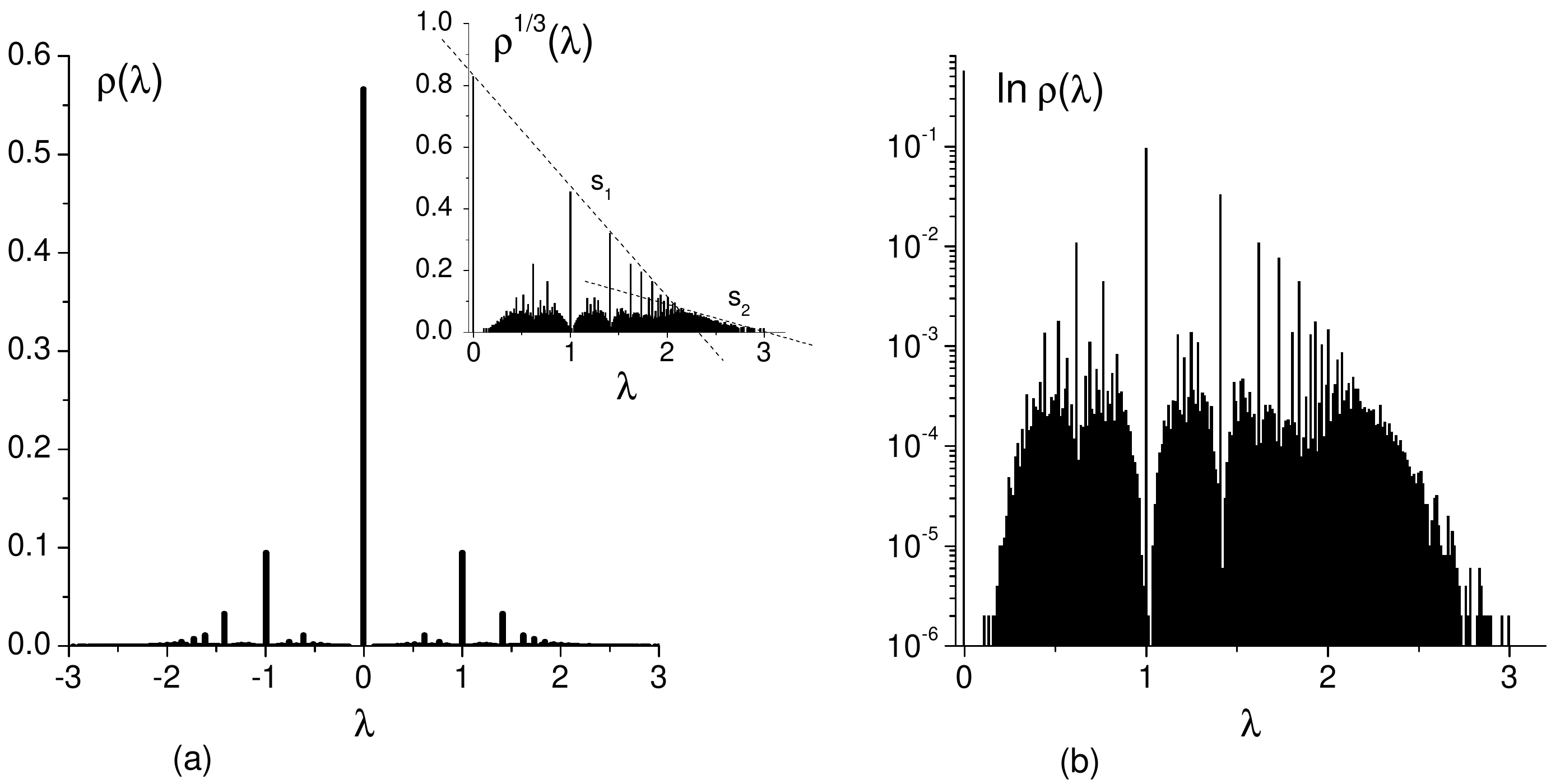}}
\caption{(a) Spectral density $\rho(\lambda)$ for ensemble of 1000 symmetric random $500\times 500$
Bernoulli matrices, generated at $q=2.0028\times 10^{-3}$ possesses regular ultrametric structure.
The fracture of the enveloping curve near $\lambda=\pm 2$ is clearly visible in the insert, where
$\rho^{1/3}(\lambda)$ is plotted; (b) the plot of $\ln \rho(\lambda)$ shows the ultrametric
structure of the spectral density and terminates close to the maximal eigenvalue for 3-branching
trees, $|\lambda_{\max}^{\rm tree}| = 2\sqrt{2}\approx 2.83$.}
\label{fig:02}
\end{figure}

The singular spectral density, $\rho(\lambda)$, for an ensemble of 1000 random symmetric matrices
$A$, each of size $500\times 500$ and generated with probability $q=2.0028\times 10^{-3}$, is shown
in \fig{fig:02}a. The spectrum possesses the regular ultrametric hierarchical structure which
becomes even more profound while plotting $\ln \rho(\lambda)$, as shown in \fig{fig:02}b. Near the
value $\lambda=\pm 2$ the enveloping curve of the function $\ln \rho(\lambda)$ exhibits a fracture,
clearly visible in the insert of \fig{fig:02}a, where the spectrum is replotted as
$\rho^{1/3}(\lambda)$ and the slops are depicted by the dashed lines $s_1$ and $s_2$. (The choice
of the exponent 1/3 for the curve $\rho^{1/3}(\lambda)$ is an experimental selection: we have just
noted that such a scaling gives the best linear shape for the enveloping curve and clearly
demonstrates the fracture of the shape near $\lambda=\pm 2$.)

The largest eigenvalue of a tree can be roughly estimated as
$|\lambda_{\max}^{\rm tree}|= 2\sqrt{p-1}$, where $p$ is the maximal vertex degree of a tree (see
e.g. \cite{rojo1,rojo2,rojo3,sen}). For linear chains, the maximal degree is $p=2$ (when the chain
length is at least three) and the maximal eigenvalue is $|\lambda_{\max}^{\rm lin}| = 2$.

In the bulk of the spectrum of sparse ensemble, the randomly branching trees (with $p=3$) and
linear chains (with $p=2$) dominate (some estimates are given in the following sections). For
$|\lambda|>|\lambda_{\max}^{\rm lin}| = 2$, the linear chains do not contribute to the spectral
density, and only graphs with $p\ge 3$ remain. For Erd\H{o}s-R\'enyi random graphs, trees with
$p\ge 4$ are rather rare, so the dominant contribution comes from graphs with $p=3$, giving
$|\lambda_{\rm max}^{\rm tree}| = 2\sqrt{2}\approx 2.83$, consistent with simulation results
(\fig{fig:02}b).

\section{Distribution of linear chains at the percolation threshold}

The spectral analysis of ensemble of random adjacency matrices $A$ at the percolation threshold is
cumbersome since the spectra of the components depend on their topology. For random
Erd\H{o}s-R\'enyi graphs, however, almost all components are trees (see e.g. \cite{Janson,book}).
Furthermore, at the percolation point (and certainly below it) the majority of subgraphs (about
95\% of them) are linear chains as we show below. The spectrum of each linear chain is easy to
compute, so to get the spectral density, $\rho_{\rm lin}(\lambda)$, of an ensemble of chain-like
subgraphs we should know the distribution, $Q_n$, of linear chains of length $n$ at the percolation
point. Below we derive an explicit expression for $Q_n$. We first recall a kinetic theory approach
which allows one to determine the size distribution of generic clusters (irrespective of their
topology) and then adopt this framework to the computation of the size distribution of linear
chains.

\subsection{General formalism: Cluster size distribution}

\begin{figure}[ht]
\centerline{\includegraphics[width=6cm]{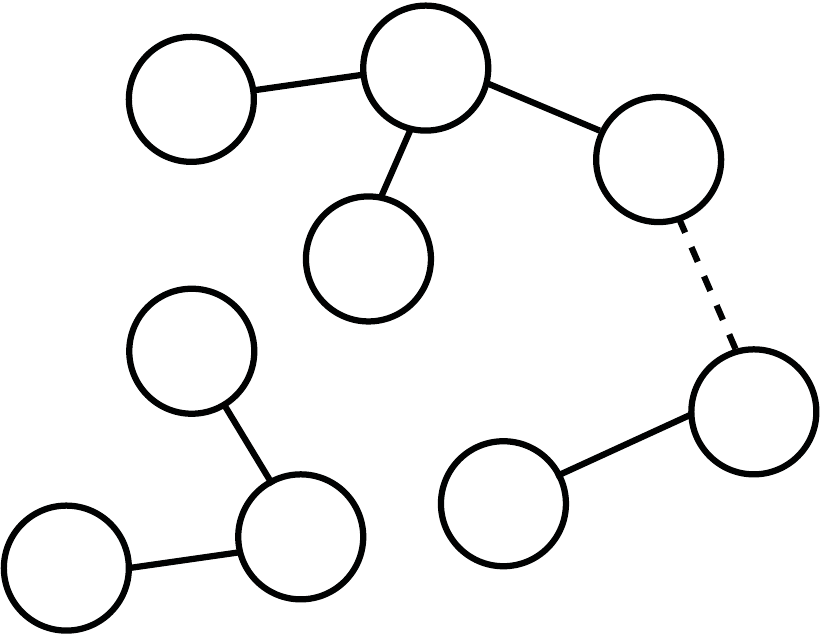}}
\caption{An evolving random graph with $N=9$. Links are indicated by solid lines. The graph is
composed of three clusters: Two clusters are chains (of length 2 and 3), the third cluster is a
star of size 4. When the link shown by a dashed line is added, the chain of length 2 and the star
merge.}
\label{fig:ill}
\end{figure}

By definition, each cluster is a maximal connected component (see Fig.~\ref{fig:ill}). To determine
the cluster size distribution we employ a dynamical framework (see e.g. \cite{book}), namely we
begin with a disjoint graph represented by a collection of $N$ separated vertices and start linking
them with a constant rate. When two vertices from distinct clusters become linked, the clusters
join. For example, the latest link in Fig.~\ref{fig:ill} joins two clusters of size $i=2$ and $j=4$
into a cluster of size $k=i+j=6$.

Generally, there are $i\times j$ ways to join distinct clusters of size $i$ and $j$. Hence, the
clusters undergo an aggregation process
\be
(i,j)\buildrel ij/(2N)\over \longrightarrow i+j
\label{agg}
\ee
which proceeds with a rate proportional to the product of their sizes. We want to determine the
cluster size distribution, $c_k(t)$. By definition, $c_k(t)=N_k(t)/N$, where $N_k(t)$ is the total
number of components with $k$ nodes. In \eqref{agg} the linking rate is chosen to be $(2N)^{-1}$.
With this choice, the cluster size distribution $c_k(t)$ satisfies
\be
\frac{dc_k}{dt}=\frac{1}{2}\sum_{i+j=k}(ic_i)(jc_j)-k\,c_k
\label{ck-eq}
\ee
The initial condition is $c_k(0)=\delta_{k,1}$. The gain term in \eq{ck-eq} accounts for clusters
generated by joining two smaller clusters whose sizes sum up to $k$. The second term on the
right-hand side of Eq.~(\ref{ck-eq}) represents loss due to linking of clusters of size $k$ to
other clusters. The corresponding gain and loss rates follow from the aggregation rule \eq{agg}.
Stopping the aggregation process at time $t$ leads to a random graph with adjacency matrix
characterized by the connectivity probability $q=t/N$.

Equations \eq{ck-eq} are well-known and they can be solved via various techniques \cite{book}. Here
we recall one approach as we shall use it in the next sub-section to determine the distribution of
linear chains. The idea is to eliminate the time dependence. Solving \eq{ck-eq} recursively one
gets
\be
c_1(t) = e^{-t}; \; c_2(t) = \tfrac{1}{2}\, t\, e^{-2t};\; c_3(t) = \tfrac{1}{2}\, t^2\, e^{-3t};\;
c_3(t) = \tfrac{2}{3}\, t^3\, e^{-4t}; \; {\rm etc}
\label{c3}
\ee
This suggests to seek the solution in the form
\be
c_k(t) = C_k\, t^{k-1}\,e^{-kt}
\label{cC}
\ee
Substituting this form into \eq{ck-eq}, we find that the coefficients $C_k$ can be obtained
recursively starting from $C_1=1$ and using
\be
(k-1)\,C_k=\frac{1}{2}\sum_{i+j=k}(iC_i)\,(jC_j)
\label{Ck-eq}
\ee
for $k>1$.  Using a generating function technique one gets $C_k=\frac{k^{k-2}}{k!}$. Thus
\be
c_k(t)=\frac{k^{k-2}}{k!}\,t^{k-1}\,e^{-kt}
\label{ckt}
\ee
The large $k$ tail is always exponential apart from the percolation point $t=t_c=1$ where
\be
c_k(1)\simeq (2\pi)^{-1/2} k^{-5/2} \qquad\text{when}\quad k\gg 1
\label{ck:crit}
\ee

\subsection{Distribution of linear chains}

Consider now linear clusters (open chains). Let $Q_k(t)$ be the density of linear clusters of
length $k\geq 2$. We have $Q_2=c_2$, and for $k\geq 3$ the chain size distribution evolves
according to
\be
\frac{dQ_k}{dt}=2c_1Q_{k-1} + 2\sum_{i+j=k} Q_i Q_j-k\,Q_k
\label{Lk-eq:long}
\ee
where the summation is taken over $i\geq 2, j\geq 2$. It is convenient to define $Q_1 =
\tfrac{1}{2}\, c_1$. This allows us to rewrite \eqref{Lk-eq:long} as
\be
\frac{dQ_k}{dt}= 2\sum_{i+j=k} Q_i Q_j-k\,Q_k
\label{Lk-eq}
\ee
here the summation runs over all $i,j$ satisfying $i+j=k$. Another virtue of the agreement $Q_1 =
\tfrac{1}{2}\, c_1$ is that \eqref{Lk-eq} is valid for all $k\geq 2$. Computing the first few
densities we get
\be
Q_1 = \tfrac{1}{2}\, c_1 = \tfrac{1}{2}\,e^{-t};\; Q_2 = c_2 = \tfrac{1}{2}\, t\,
e^{-2t}; \; Q_3 = c_3 = \tfrac{1}{2}\, t^2\, e^{-3t}; \; Q_3 =  \tfrac{1}{2}\, t^3\, e^{-4t}
\label{L4}
\ee
\emph{etc} (we have used \eqref{Lk-eq} to determine $Q_4$). One is led to the conjectural behavior
\be
Q_k(t) = \tfrac{1}{2}\, t^{k-1}\,e^{-kt}
\label{LL}
\ee
This indeed provides the solution to \eqref{Lk-eq}  and the initial condition
$Q_k(0)=\tfrac{1}{2}\, \delta_{k,1}$.

Thus, the linear chain length distribution \eqref{LL} is {\em purely} exponential. Rewriting
\eqref{LL} as
\be
Q_k(t) = \frac{1}{2t}\, e^{-k(t-\ln t)}
\label{polymers}
\ee
we see that the least-steepest decay occurs at the percolation point $t=t_c=1$ where
\be
Q_k\equiv Q_k(1) = \frac{1}{2}\, e^{-k}
\label{polymers:crit}
\ee

The total \emph{chain density} at the percolation point is
\be
c_1+\sum_{j\geq 2}Q_j(1) = \frac{1}{2}\,\frac{2-e^{-1}}{e-1}=0.474928\ldots
\label{density:crit}
\ee
Comparing this result with the total cluster density at the percolation point, $\sum_{k\geq
1}c_k(1) = \frac{1}{2}$, we conclude that at the percolation point almost 95\% of clusters are
linear chains. This justifies the abundance of linear chains (see \fig{fig:01}).

Note that the total \emph{mass density} of linear chains at the percolation point,
\be
c_1+\sum_{j\geq 2} jQ_j(1) = \frac{1}{2 e}\left[1+(1-e^{-1})^{-2}\right]=0.6442765\ldots,
\label{mass:crit}
\ee
is significantly smaller than unity. Large clusters are usually not linear and hence despite of
their rareness they provide a notable contribution to the mass density.

\subsection{Comments on more complicated components and different random graph models}

Below the percolation point, the total number of components which are not trees remains finite in
the thermodynamic limit $N\to\infty$, so it greatly fluctuates from realization to realization. The
components which are not trees are \emph{unicyclic}: each such component contains one loop. At the
percolation point the total number of unicyclic components diverges but very slowly, viz. as
$\frac{1}{6}\ln N$. These results were established using probabilistic techniques, see
\cite{Janson}; they are also easily derivable using above dynamical approach (see \cite{bk04}).
Overall we see that the contribution of the non-trees is negligible.

The influence of the "shape" of the tree on the spectra is more notable. We have shown, however,
that the majority of the clusters are not merely trees, but linear chains. In principle, one can
modify the definition \eqref{eq:01} to ensure that all components are chains. If each row (and
column) of the adjacency matrix $A$ has no more than two non-zero elements, $\sum_{j=1}^N a_{ij}\le
2$ for $i=1,...,N$, the corresponding graph is obviously the collection of linear chains (including
loops) of various lengths.

Equivalently, the chain model can be defined dynamically by using the same procedure as before and
checking that the constraint is obeyed. In other words, we attempt to link vertices randomly, but
we actually draw a link only if the degrees of both vertices are smaller than 2. The cluster size
distribution characterizing the chain model evolves according to master equation
\be
\frac{dc_k}{dt}=\frac{1}{2}\sum_{i+j=k} K_{ij} c_i c_j-c_k\sum_{j\geq 1} K_{kj} c_j
\label{ck-linear}
\ee
with reaction rate matrix
\be
K_{ij} = \left(\begin{array}{ccccc}
1 & 2 & 2 & 2 & \cdots \\
2 & 4 & 4 & 4 &   \\
2 & 4 & 4 & 4 &   \\
2 & 4 & 4 & 4 &   \\
\vdots &  & &  & \ddots
\end{array} \right)
\label{Kij}
\ee
The initial condition is $c_k(0)=\delta_{k,1}$. A more general model with reaction rate matrix of
the type \eqref{Kij} composed of three arbitrary rates has been studied in \cite{Mauro}.

Using \eqref{ck-linear}--\eqref{Kij} one finds that the density of isolated vertices $c_1(t)$ and
the total cluster density $c(t)=\sum_{j\geq 1} c_j(t)$ satisfy a pair of coupled rate equations:
\be
\frac{dc_1}{dt}=-c_1(2c-c_1), \quad  \frac{dc}{dt}=-\tfrac{1}{2}(2c-c_1)^2
\label{cc-linear}
\ee
The initial condition becomes $c(0)=c_1(0)=1$. Treating $c$ as a function of $c_1$ we recast
\eqref{cc-linear} into $\frac{dc}{dc_1}=\frac{c}{c_1}-\frac{1}{2}$, from which
\be
c=c_1\left[1-\tfrac{1}{2}\ln c_1\right]
\label{cc-sol}
\ee
Plugging this into the first equation \eqref{cc-linear} we obtain an implicit solution
\be
\int_{c_1}^1 \frac{du}{u^2(1-\ln u)} = t
\label{c1-sol}
\ee
The following densities can be found by re-writing the rate equations \eqref{ck-linear} in a
manifestly recursive form,
\begin{equation*}
\begin{split}
\frac{dc_2}{dt} &= \tfrac{1}{2}c_1^2 - 2c_2(2c-c_1),\\
\frac{dc_3}{dt} &= 2c_1 c_2 - 2c_3(2c-c_1),\\
\frac{dc_4}{dt} &= 2c_1 c_3 + 2c_2^2 - 2c_4(2c-c_1),\\
\frac{dc_5}{dt} &= 2c_1 c_4 + 4c_2 c_3 - 2c_5(2c-c_1),
\end{split}
\end{equation*}
etc., and solving them recursively.

We use the notation $c_k$, but we emphasize that clusters generated by the above procedure are
linear chains. There is no percolation transition in this model, so when should we stop the
process? Recall that for Erd\H{o}s-R\'enyi random graphs, the percolation occurs when the average
degree is equal to one. In the model \eqref{ck-linear}--\eqref{Kij}, the fraction of vertices of
degree zero ($d_0$), vertices of degree one ($d_1$), and vertices of degree two ($d_2$) are
\begin{equation*}
\begin{split}
d_0 &= c_1\\
d_1 &= 2\sum_{j\geq 2} c_j = 2(c-c_1)\\
d_2 &= 1-d_0-d_1 = 1-c_1 - 2(c-c_1)
\end{split}
\end{equation*}
The average degree $\langle d\rangle = d_1 + 2d_2$ is therefore  $\langle d\rangle = 2(1-c)$. Thus,
in analogy with Erd\H{o}s-R\'enyi random graphs, we stop the process when the average degree
reaches unity, the corresponding time $t_s$ is found from $c(t_s)=\frac{1}{2}$. Another choice is
to run the process till the very end when the system turns into a collection of closed loops, see
\cite{bk:ring} for this and other models of ring formation.

One can generalize above model by requiring that each row (and column) of the adjacency matrix $A$
has no more than $p$ non-zero elements: $\sum_{j=1}^N a_{ij}\le p$ for $i=1,...,N$. This model can
be investigated in the dynamic framework. Unfortunately, it has proved to be much more difficult
for the analysis \cite{bk:regular} than the extreme versions corresponding to the chain model
($p=2$) and the Erd\H{o}s-R\'enyi variant ($p=N-1$).

\section{Spectral density of canonical ensemble of bi-diagonal random matrices}

\subsection{General formalism: Spectral density}

We have seen in the Section III that the fraction of linear clusters of various lengths is about
95\% at the percolation threshold in the ensemble of all sparse graphs. This justifies our study of
ensembles of linear clusters (open chain and loops) only.

It is convenient to represent the linear cluster ensemble at the percolation point (with the
distribution $Q_k$ in chain lengths) by the bi-diagonal $N\times N$ symmetric matrix $B$, where
\be
B = \left(\begin{array}{ccccc}
0 & x_1 & 0 & 0 & \cdots \\  x_1 & 0 & x_2 & 0 &  \\  0 & x_2 & 0 & x_3 &  \\
0 & 0 & x_3 & 0 &  \\ \vdots &  &  &  & \ddots
\end{array} \right)
\label{eq:06}
\ee
and the distribution of each $x_i$ ($i=1,...,N$) is Poissonian:
\be
x_i=\left\{\begin{array}{ll} 1 & \mbox{with probability $q$} \medskip \\
0 & \mbox{with probability $1-q$} \end{array} \right.
\label{eq:06a}
\ee
Due to ergodicity, the spectral density, $\rho_{\rm lin}(\lambda)$, of the ensemble of $N\times N$
matrices $B$ in the limit $N\to\infty$ coincides with the spectral density of an individual matrix
$B$ and can be computed straightforwardly. Note that for any $x_k=0$, the matrix $B$ splits into
independent fragments. So, it is instructive to consider subchains consisting of sequential set of
$x_k=1$. The probability to have a subchain with $n$ consecutive 1 is $\tilde{Q}_n=q^n$, as it
follows from \eq{eq:06a}. Comparing $\tilde{Q}_n$ with the distribution $Q_n=\frac{1}{2}e^{-n}$ of
linear subchains in ultra-sparse matrix ensemble at the percolation point (see \eq{polymers:crit}),
we conclude that $q=e^{-1}$.

In what follows we derive $\rho_{\rm lin}(\lambda)$ for arbitrary values of $q$. This allows us to
investigate the spectral density of the ensemble of bi-diagonal random operators in the limit $q\to
1$ and uncover it number-theoretic structure.

The spectrum of linear chain of length $n$ (i.e. of symmetric $n\times n$ bi-diagonal matrix $B$
with all $x_k=1$, $k=1,...,n$) reads
\be
\lambda_k = 2\cos\frac{\pi k}{n+1}; \qquad (k=1,...,n)
\label{eq:06b}
\ee
The spectral density of the ensemble of $N\times N$ random matrices $B$ with the Poissonian
distribution of matrix elements can be written as
\begin{eqnarray}
\rho_{\rm lin}(\lambda) &=& \lim_{N\to\infty}\frac{1}{N}\la \sum_{k=1}^{N} \delta(\lambda-\lambda_k)
\ra \nonumber\\
&=& \lim_{N\to\infty \atop \eps\to 0} \frac{\eps}{\pi N} {\rm Im}\, \la G_N(\lambda - i\eps) \ra
\nonumber\\
&=& \lim_{N\to\infty \atop \eps\to 0} \frac{\eps}{\pi N} \sum_{n=1}^N \tilde{Q}_n\; {\rm Im}\,
G_n(\lambda - i\eps)
\label{eq:08}
\end{eqnarray}
where we have used an identity
$$
\lim_{\eps\to+0} \im \frac{1}{\lambda\pm i \eps} = \mp i \pi \delta(\lambda)
$$
The Green function
\be
G_n(\lambda-i\eps) = \sum_{k=1}^n \frac{1}{\lambda-\lambda_k-i\eps}
\label{eq:09}
\ee
with $\lambda_k$ given by \eq{eq:06b}, is associated with each particular gapless matrix $B$ of $n$
sequential ones on the sub-diagonals, while $\la ...\ra$ means averaging over the distribution
$\tilde{Q}_n=q^n$.

Collecting \eq{eq:06b}, \eq{eq:08} and \eq{eq:09}, we find an explicit expression for the density
of states, $\rho_{\rm lin}(\lambda)$:
\be
\rho_{\rm lin}(\lambda) = \lim_{N\to\infty \atop \eps\to 0} \frac{\eps}{\pi N} \sum_{n=1}^{N} q^n
\sum_{k=1}^n\frac{1}{\left(\lambda-2\cos\frac{\pi k}{n+1}\right)^2+\eps^2}
\label{eq:10}
\ee

\begin{figure}[ht]
\centerline{\includegraphics[width=15cm]{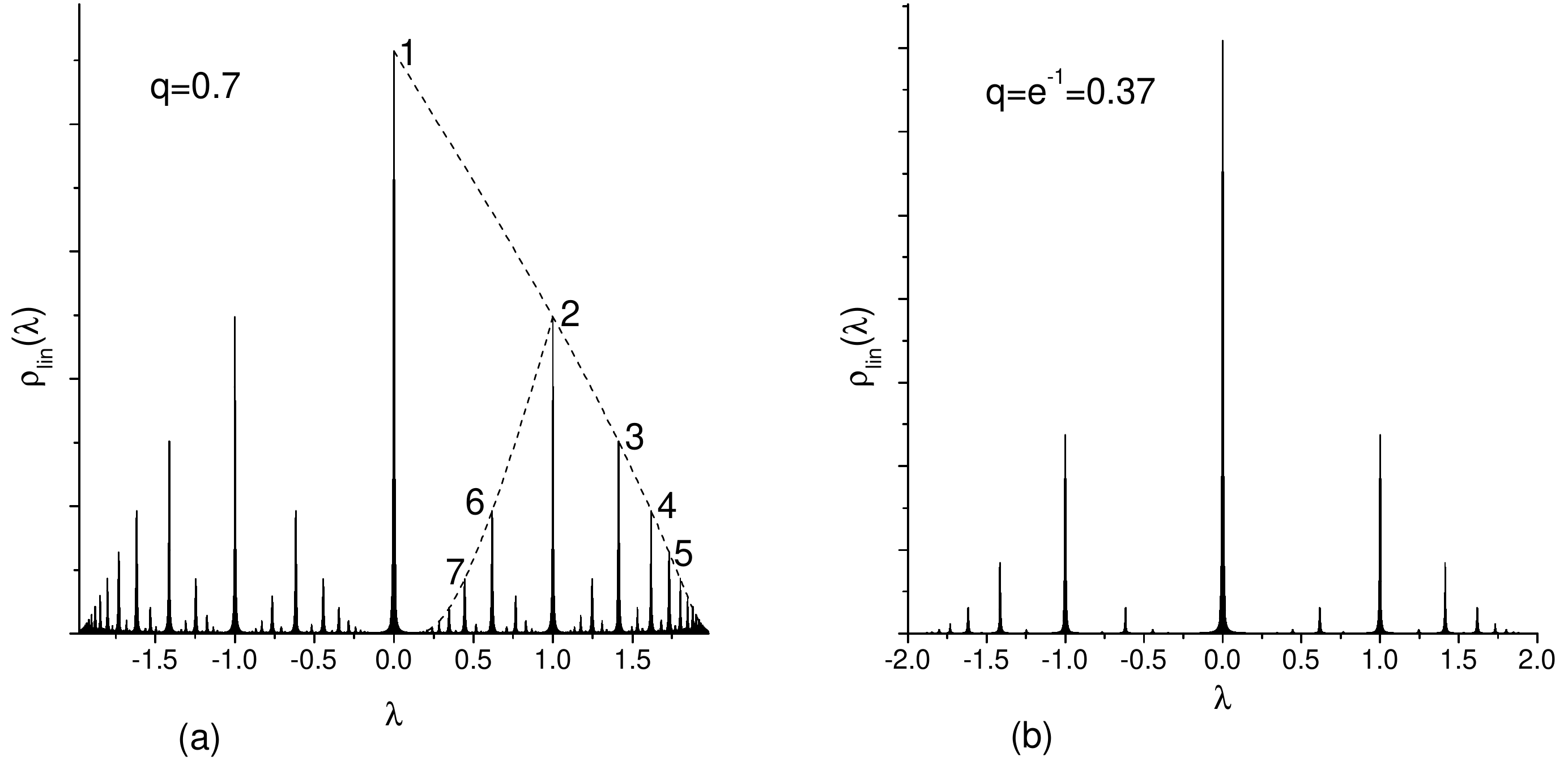}}
\caption{The spectral density $\rho_{\rm lin}(\lambda)$ for the ensemble of bi-diagonal matrices
at $q=0.7$ (a) and $q=e^{-1}\approx 0.37$ (b); the regularization parameter $\eps$ is taken
$\eps=2\times 10^{-3}$.}
\label{fig:03}
\end{figure}

The sample plots $\rho_{\rm lin}(\lambda)$ for two different values of $q$, namely for $q=0.7$ and
$q=e^{-1}\approx 0.37$ computed numerically via \eq{eq:10} with $\eps=3\times 10^{-3}$ are shown in
the \fig{fig:03}.

\subsection{Analysis of enveloping curves and tails of spectral density $\rho_{\rm lin}(\lambda)$}

One can compute the enveloping curves for any monotonic sequence of peaks. In \fig{fig:03}, we show
two series of sequential peaks: $S_1=\{$1--2--3--4--5--...$\}$ and $S_2=\{$2--6--7--...$\}$. Any
monotonic sequence of peaks corresponds to the set of eigenvalues $\lambda_k$ constructed on the
basis of Farey sequence \cite{farey}. For example, as shown below, the peaks in the series $S_1$
are located at $\lambda_k=-2\cos\frac{\pi k}{k+1}$, ($k=1,2,...$), while the peaks in the series
$S_2$ are located at $\lambda_{k'} = -2\cos \frac{\pi k'}{2k'-1}$, ($k'=2,3,...$). Positions of
peaks obey the following rule: let $\{\lambda_{k-1},\, \lambda_k,\, \lambda_{k+1}\}$ be three
consecutive monotonically ordered peaks (e.g., peaks 2--3--4 in \fig{fig:03}), and let
$$
\lambda_{k-1}=-2\cos \frac{\pi p_{k-1}}{q_{k-1}}, \quad\lambda_{k+1}=-2\cos \frac{\pi
p_{k+1}}{q_{k+1}}
$$
where $p_k$ and $q_k$ ($k=1,...,N$) are coprimes. The position of the intermediate peak,
$\lambda_k$, is
\be
\lambda_{k}=-2\cos \frac{\pi p_{k}}{q_{k}}\,, \qquad
\frac{p_{k}}{q_{k}} = \frac{p_{k-1}}{q_{k-1}} \oplus \frac{p_{k+1}}{q_{k+1}}
\equiv \frac{p_{k-1}+p_{k+1}}{q_{k-1}+q_{k+1}}
\label{eq:11}
\ee

\begin{figure}[ht]
\centerline{\includegraphics[width=15cm]{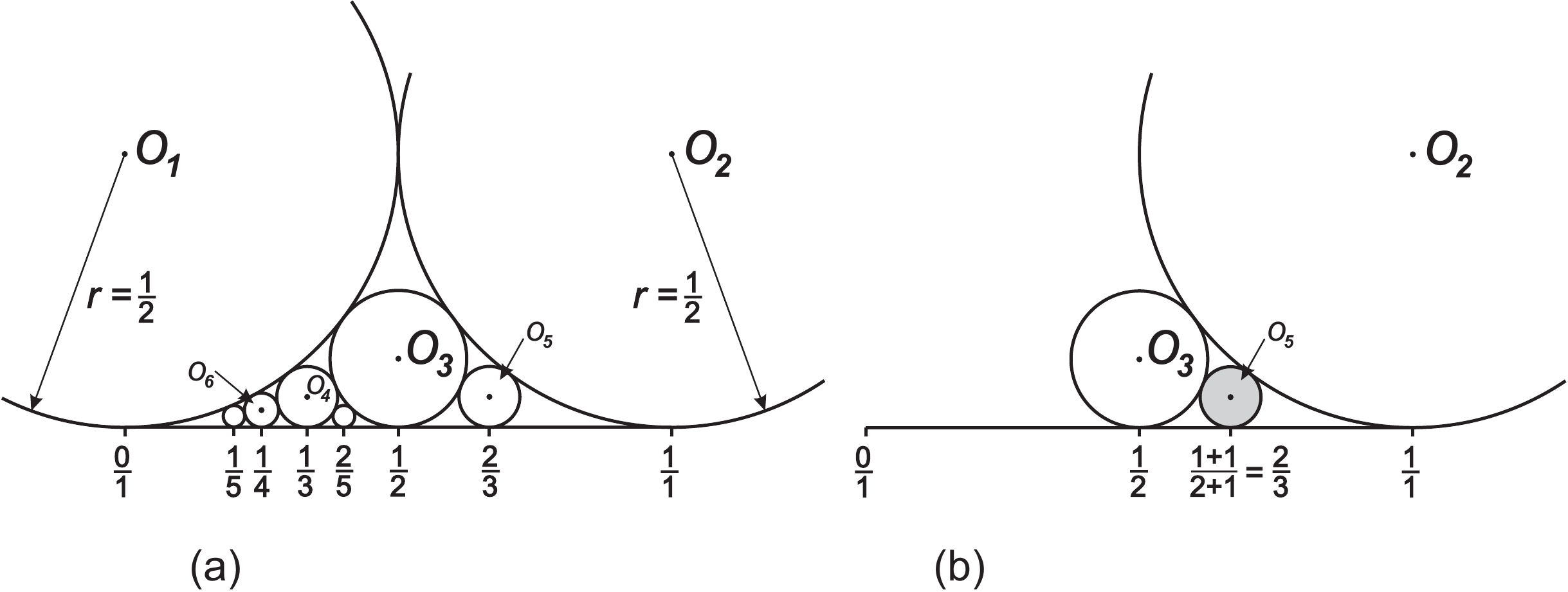}}
\caption{Ford circles as illustration of the Farey sequence construction: a) Each circle touches two
neighbors (right and left) and the segment; b) Determination of the position of newly generated
circle via the $\oplus$ addition: $\frac{p_{k-1}}{q_{k-1}} \oplus
\frac{p_{k+1}}{q_{k+1}}=\frac{p_{k-1}+p_{k+1}}{q_{k-1}+q_{k+1}}$.}
\label{fig:farey}
\end{figure}

The sequences of coprime fractions constructed via the $\oplus$ addition are known as Farey
sequences. A simple geometric model behind the Farey sequence, known as Ford circles
\cite{ford}, is shown in \fig{fig:farey}a. Take the segment $[0,1]$ and draw two circles $O_1$ and
$O_2$ both of radius $r=\frac{1}{2}$, which touch each other, and the segment at the points 0 and
1. Now inscribe a new circle $O_3$ touching $O_1$, $O_2$ and $[0,1]$. Where is the position of the
new circle along the segment? The generic recursive algorithm is shown in \fig{fig:farey}b and
constitutes the Farey sequence construction.

Consider two examples related to the set of eigenvalues.
\begin{itemize}
\item The monotonic series $S_1$ is can be constructed recursively by taking
$$
\lambda_1\equiv -2\cos\frac{\pi}{2}=0, \quad \lambda_{\infty} \equiv -2\cos\frac{\pi}{1}=2
$$
The next-to-highest peak 2 is located at
$$
\lambda_2=-2\cos \left[\pi\left(\frac{1}{2} \oplus \frac{1}{1}\right)\right]=-2\cos\frac{2\pi}{3}=1
$$
Peak 3 is located at
$$
\lambda_3=-2\cos \left[\pi\left(\frac{2}{3} \oplus
\frac{1}{1}\right)\right]=-2\cos\frac{3\pi}{4}=\sqrt{2}
$$
Peak 4 is located at
$$
\lambda_4=-2\cos \left[\pi\left(\frac{3}{4} \oplus
\frac{1}{1}\right)\right]=-2\cos\frac{4\pi}{5}=\frac{\sqrt{5}+1}{2}
$$
etc. The generic expression for $\lambda_k$ in the series $S_1$ reads
\be
\lambda_k =-2\cos\frac{k\pi}{k+1}, \quad (k=1,2,...)
\label{eq:s1}
\ee

\item The monotonic series $S_2$ is can be constructed similarly by taking as reference points
$\lambda_1$ and $\lambda_2$. For peak 6 we have
$$
\lambda_6=-2\cos \left[\pi\left(\frac{1}{2} \oplus \frac{2}{3}\right)\right]= -2\cos\frac{3\pi}{5}
= \frac{\sqrt{5}-1}{2}
$$
Peak 7 is located at
$$
\lambda_7=-2\cos \left[\pi\left(\frac{1}{2} \oplus \frac{3}{5}\right)\right]= -2\cos\frac{4\pi}{7}
$$
etc. The generic expression for $\lambda_{k'}$ in the series $S_2$ reads
\be
\lambda_{k'} =-2\cos\frac{k'\pi}{2k'-1}, \quad (k'=2,3,...)
\label{eq:s2}
\ee
\end{itemize}

\begin{figure}[ht]
\centerline{\includegraphics[width=15cm]{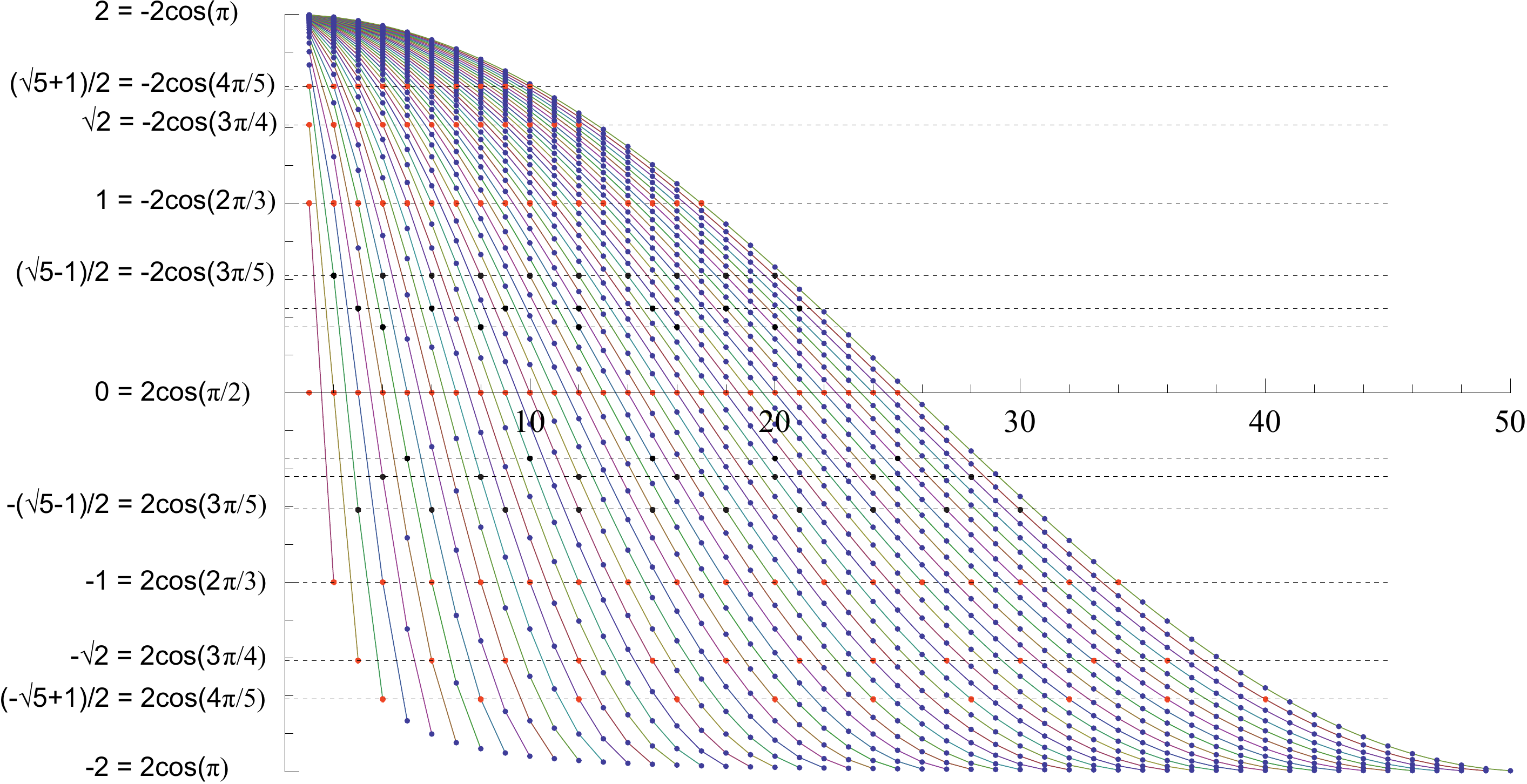}}
\caption{Family of 50 curves $w_n(k)=-2\cos\frac{\pi k}{n+1}$. Each curve corresponds to a
particular $n=1,...,50$, the points mark values $k=1,...,n$. Each horizontal line is the multiplicity
of the corresponding root and contributes to the height of the peak in the spectral density.}
\label{fig:04}
\end{figure}

As follows from \eq{eq:10}, the function $\rho_{\rm lin}(\lambda)$ is nonzero only at $\lambda_k$.
In \fig{fig:04} we show the set of eigenvalues $\lambda_n(k)=-2\cos\frac{\pi k}{n+1}$
for $n=1,..., 50$. Every point designates some eigenvalue, $\lambda_k$ (see \eq{eq:06b}); the
points along each curve correspond to \emph{different} values $k$ for a \emph{specific} value $n$.
The collection of eigenvalues in each horizontal line, coming from different $n$, gives the
multiplicity of corresponding eigenvalue $\lambda_k$ and contributes to the height of a peak in the
spectral density $\rho_{\rm lin}(\lambda)$.

The heights of peaks in any monotonic sequence (like $S_1$ or $S_2$) can be computed directly from
\eq{eq:10} by summing geometric series along each horizontal line in \fig{fig:04}. For example,
for the sequences $S_1$ and $S_2$ we have:
\be
\mbox{$S_1$}: \left\{\begin{array}{ll} \lambda_1=-2\cos\frac{\pi}{2} & \quad \rho^{S_1}(\lambda_1)
= q^1+q^3+q^5+q^{9}+... \medskip \\ \lambda_2=-2\cos\frac{2\pi}{3} & \quad \rho^{S_1}(\lambda_2) =
q^2+q^5+q^8+q^{11}+...
\medskip \\ & ... \\ \lambda_k=-2\cos\frac{\pi k}{k+1} & \quad \rho^{S_1}(\lambda_k) =
\sum\limits_{s=1}^{\infty} q^{(k+1)s-1} = \frac{q^k}{1-q^{k+1}} \quad (k=1,2,...)
\end{array} \right.
\label{eq:12a}
\ee
and
\be
\mbox{$S_2$}: \left\{\begin{array}{ll} \lambda_2=-2\cos\frac{2\pi}{3} & \quad \rho^{S_2}(\lambda_2)
= q^2+q^5+q^8+q^{11}+... \medskip \\ \lambda_3=-2\cos\frac{3\pi}{5} & \quad \rho^{S_2}(\lambda_3) =
q^4+q^9+q^{14}+q^{19}+...
\medskip \\ & ... \\ \lambda_{k'}=-2\cos\frac{\pi k'}{2k'-1} & \quad \rho^{S_2}(\lambda_{k'}) =
\sum\limits_{s=1}^{\infty} q^{(2k'-1)s-1} = \frac{q^{2k'-2}}{1-q^{2k'-1}} \quad (k=2,3,...)
\end{array} \right.
\label{eq:12b}
\ee

\begin{figure}[ht]
\centerline{\includegraphics[width=14cm]{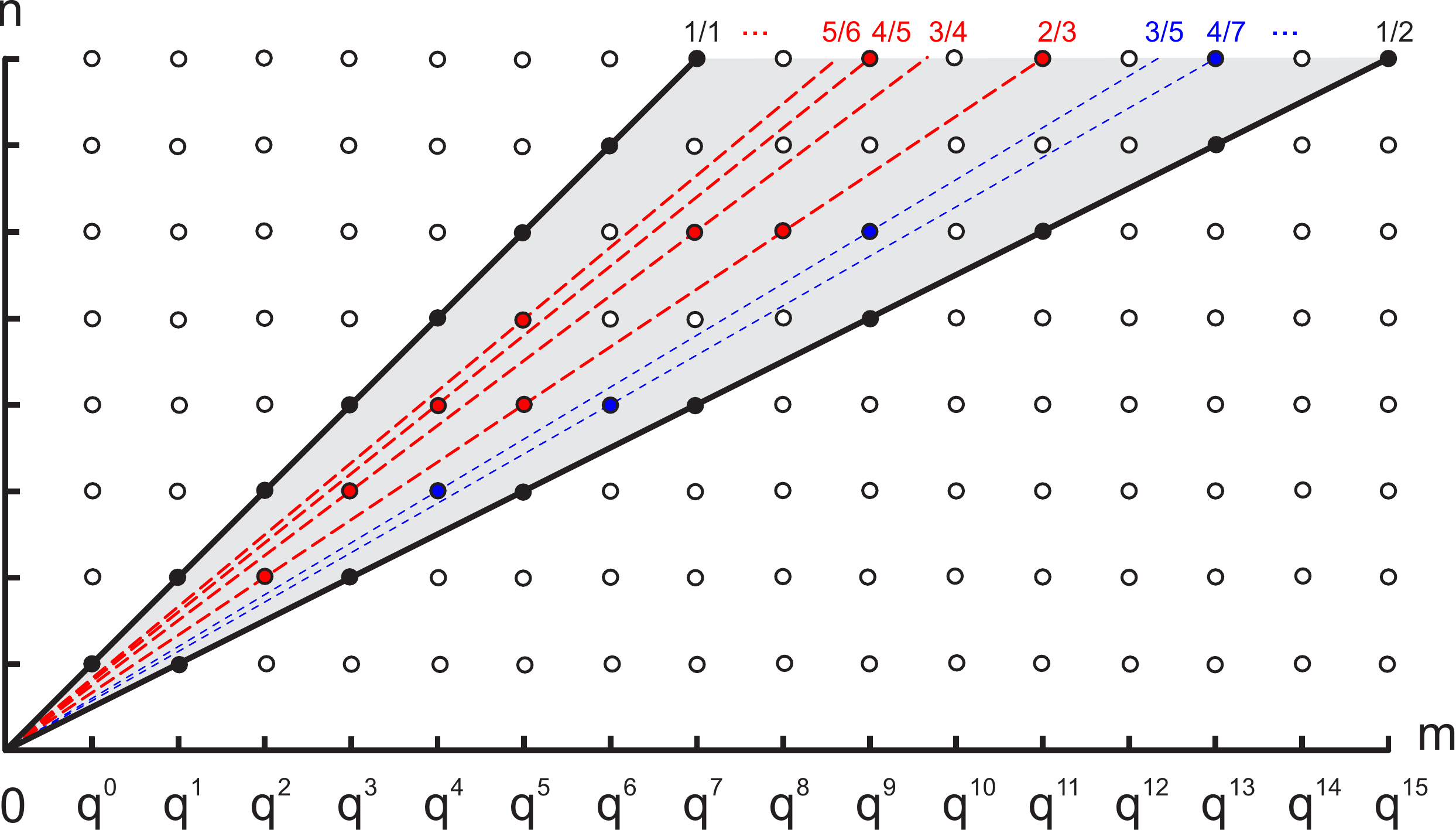}}
\caption{Visibility diagram. Each point in the vertical row $m$ carries a weight $q^m$.
Integer points within the wedge $[\pi/4,\pi/2]$ are designated by circles. The dashed rays are
drawn at rational angles $\alpha_{p,q} = \arctan \frac{p}{q}$. The weights $q^m$ corresponding to
marked integer points, are summed up along the rays.}
\label{fig:05}
\end{figure}

Note that the expressions \eq{eq:12a}--\eq{eq:12b} are tightly linked to so-called visibility
diagram (our notion is slightly modified with respect to a visibility diagram defined in
\cite{vallet}). Consider the square lattice of integer points $(m,n)$ and add a weight $q^m$ to
each vertical row. Emit rays at rational angles $\alpha_{p,q}=\arctan \frac{p}{q}$ from the point
$(0,0)$ within the wedge $[\pi/4,\pi/2]$ in the positive direction as shown in \fig{fig:05}.

Let us sum up the weights, $q^m$, of integer points along each ray emitted at the rational slope
$\frac{p}{q}$. For example, consider the sequence of slopes $\tan \alpha_{1,k}=\frac{1}{k}$ with
$k=1,2,...$ and $\tan \alpha_{k',2k'-1}=\frac{k'}{2k'-1}$ shown in \fig{fig:05}. One can verify
that these geometric series exactly reproduce the sequences $S_1$ and $S_2$ in
Eqs.~\eq{eq:12a}--\eq{eq:12b} at corresponding $k$ and $k'$.

Two parametrically defined enveloping curves, namely $\left(\lambda_k, \rho^{S_1}(k)\right)$ for the
sequence $S_1$ $(k=1,2,...)$ and $\left(\lambda_{k'}, \rho^{S_2}(k')\right)$ for the sequence $S_2$
$(k'=2,3,...)$, are shown in \fig{fig:03}a by dashed lines. Equations \eq{eq:12a}--\eq{eq:12b}
allow to find the singular tails of distributions $\rho^{S_1}(\lambda)$ as $\lambda\to 2$ and
$\rho^{S_2}(\lambda)$ as $\lambda\to 0$. Expand now the parametrically defined curves in
\eq{eq:12a}--\eq{eq:12b} at $k\to\infty$ and $k'\to\infty$:
\be
\begin{array}{lll}
\disp S_1: & \lambda=-2\cos \frac{\pi k}{k+1}\Big|_{k\to\infty}\to 2-\frac{\pi^2}{k^2}, &
\disp \quad \rho^{S_1}(\lambda_k)\Big|_{k\to\infty}\to q^{k} \medskip \\
\disp S_2: & \lambda'=-2\cos\frac{\pi k'}{2k'-1}\Big|_{k'\to \infty}\to \frac{\pi}{2k'}, & \disp
\quad \rho^{S_2}(\lambda_{k'})\Big|_{k'\to \infty} \to q^{2k'}
\end{array}
\label{eq:tails}
\ee
Expressing $k$ and $k'$ in \eq{eq:tails} in terms of $\lambda$ and $\lambda'$, we get
\be
k\to \frac{\pi}{\sqrt{2-\lambda}}, \quad k'\to \frac{\pi}{2\lambda'}
\label{eq:tails2}
\ee
Asymptotics \eq{eq:tails2} together with \eq{eq:tails} permit us to find the singular tails of the
eigenvalue distribution $\rho_{\rm lin}(\lambda)$ at $\lambda \to 2$ and $\lambda'\to 0$:
\be
\rho^{S_1}_{\rm lin}(\lambda\to 2) \to q^{\pi/\sqrt{2-\lambda}}, \quad \rho^{S_2}_{\rm
lin}(\lambda'\to 0) \to q^{\pi/\lambda'}
\label{eq:anderson}
\ee

The edge singularity $\rho^{S_1}_{\rm lin}(\lambda\to 2)$ at $\lambda\to 2$ in Eq.~\eq{eq:anderson}
of the eigenvalue distribution reproduces the corresponding Lifshitz tail of the one-dimensional
Anderson localization in random Schr\"odinger operator \cite{pastur,kirsch}: $\rho(E)\sim
e^{-E^{-d/2}}$, where $E=2-\lambda$ and $d=1$. The appearance of the edge singularity
$\rho(E)\sim e^{-1/\sqrt{E}}$ in our situation is purely geometric, it is not relied
on any entropy-energy-balance consideration like optimal fluctuation \cite{pastur}.

\section{Spectral density of bi-diagonal matrices in the limit $q\to 1$ and the Dedekind
$\eta$-function}

\subsection{Limiting properties of the Dedekind $\eta$-function via duality relation}

Recall the definition of the Dedekind $\eta$-function \cite{chand}:
\be
\eta(z)=e^{\pi i z/12}\prod_{n=0}^{\infty}(1-e^{2\pi i n z})\,; \quad z=x+i y \quad (y>0)
\label{eq:14}
\ee
Consider now the following function
\be
f(z) = A^{-1}\, |\eta(z)| (\im z)^{1/4}; \qquad
A=\left|\eta\left(\frac{1}{2}+i\frac{\sqrt{3}}{2}\right)
\right|\left(\frac{\sqrt{3}}{2}\right)^{1/4} \approx 0.77230184...
\label{eq:15}
\ee
This function satisfies the following duality relation
\be
f\left(\left\{\frac{m}{k}\right\} + iy\right) = f\left(\left\{\frac{s}{k}\right\}+\frac{i}{k^2y}
\right); \quad ms-kr = 1; \quad \{m,s,k,r\}\in \mathbb{Z}; \quad y>0
\label{eq:15a}
\ee
where $\left\{\frac{m}{k}\right\}, \left\{\frac{s}{k}\right\}$ denote fractional parts of
corresponding quotients. Two particular identities follow from \eq{eq:15a}:
\be
\begin{split}
f\left(\frac{k}{k+1} + iy\right) &= f\left(\frac{1}{k+1}+\frac{i}{(k+1)^2y} \right)\\
f\left(\frac{k'}{2k'-1} + iy\right) &= f\left(\frac{2}{2k'-1}+ \frac{i}{(2k'-1)^2y} \right)
\end{split}
\label{eq:16}
\ee

Using \eq{eq:16} we can obtain the asymptotics of $\eta(z)$ when $y\to 0^+$. Take into account the
relation of Dedekind $\eta$--function with Jacobi elliptic functions:
\be
\vartheta_1'(0,e^{\pi i z})=\eta^3(z)
\label{eq:17}
\ee
where
\be
\vartheta_1'(0,e^{\pi i z})\equiv \frac{d\vartheta_1(u,e^{\pi i z})}{du}\bigg|_{u=0}= e^{\pi i
z/4}\sum_{n=0}^{\infty}(-1)^n (2n+1)e^{\pi i n(n+1)z}
\label{eq:18}
\ee
Rewrite \eq{eq:16} as follows:
\be
\disp \left|\eta\left(\frac{k}{k+1} + iy\right)\right| = \frac{1}{\sqrt{(k+1)y}}
\left|\eta\left(\frac{1}{k+1}+\frac{i}{(k+1)^2y} \right)\right|
\label{eq:19}
\ee
(Since the computations are identical for both equations in \eq{eq:16}, we do not reproduce the
derivation for the second one, and will write for it the final answer only). Applying
\eq{eq:17}--\eq{eq:18} to \eq{eq:19}, we get:
\be
\left|\eta\left(\frac{k}{k+1} + iy\right)\right|=\frac{1}{\sqrt{(k+1)y}}\left|e^{\frac{\pi}{4} i
\left(\frac{1}{k+1}+\frac{i}{(k+1)^2y}\right)}\sum_{n=0}^{\infty}(-1)^n (2n+1)e^{\pi i
n(n+1)\left(\frac{1}{k+1}+\frac{i}{(k+1)^2y} \right)} \right|^{1/3}
\label{eq:20}
\ee
Equation \eq{eq:20} enables us to extract the leading asymptotics of $\left|\eta\left(\frac{k}{k+1}
+ iy\right)\right|$ in the $y\to 0^+$ limit. Noting that every term of the series in \eq{eq:20} for
any $n\ge 1$ and for small $y$ (i.e. for $y\ll (k+1)^{-2}$) converges exponentially fast, we have
\be
\left|\eta\left(\frac{k}{k+1} + iy\right)\right|_{y\to 0^+}\to\frac{1}{\sqrt{(k+1)y}}
e^{-\frac{\pi}{12 (k+1)^2y}};
\label{eq:21}
\ee
Similarly, we get the asymptotics for $\left|\eta\left(\frac{k'}{2k'-1} + iy\right)\right|$ at
$y\to 0^+$ (valid for $y\ll (2k'-1)^{-2}$). Thus, we have finally:
\be
\begin{array}{lcl}
\disp \sqrt{-\ln \left|\eta\left(\frac{k}{k+1} + iy\right)\right|}\Bigg|_{y\to 0^+} & \to & \disp
\frac{1}{k+1}\sqrt{\frac{\pi}{12y}} \medskip \\ \disp \sqrt{-\ln \left|\eta\left(\frac{k'}{2k'-1} +
iy\right)\right|}\Bigg|_{y\to 0^+} & \to & \disp \frac{1}{2k'-1}\sqrt{\frac{\pi}{12y}}
\end{array}
\label{eq:23}
\ee
In \eq{eq:23} we have dropped out the subleading terms at $y\to 0^+$, which are of order of $O(\ln
y)$.

\subsection{Tails of bi-diagonal matrices at $q\to 1$ and relation with Dedekind $\eta$-function}

Expressions \eq{eq:12a}--\eq{eq:12b} allow to analyze the tails of the distribution $\rho_{\rm
lin}(\lambda)$ when $q\to 1$ and $k$ is fixed. Namely, taking the limit $q\to 1$ in
\eq{eq:12a}--\eq{eq:12b}, we arrive at:
\be
\begin{array}{l}
\disp \rho_1^{S_1}(\lambda_k)\big|_{q\to 1} = \left.\frac{q^k}{1-q^{k+1}} \right|_{q\to 1} \to
\frac{1}{(k+1)(1-q)} \medskip \\
\disp \rho_1^{S_2}(\lambda_{k'})\big|_{q\to 1} = \left. \frac{q^{2k'-2}}{1-q^{2k'-1}} \right|_{q\to
1} \to \frac{1}{(2k'-1)(1-q)}
\end{array}
\label{eq:13a}
\ee
These expressions lead us to the following asymptotics of tails of the distribution $\rho(\lambda)$
at $\lambda\approx 2-\frac{\pi^2}{k^2}\big|_{k\gg 1} \to 2$ and $\lambda' \approx
\frac{\pi}{2k'}\big|_{k\gg 1}\to 0$:
\be
\begin{array}{l}
\disp \rho_1^{S_1}\big|_{q\to 1}(\lambda\to 2) \to  \frac{\sqrt{2-\lambda}}{\pi(1-q)} \medskip \\
\disp \rho_1^{S_2}\big|_{q\to 1}(\lambda'\to 0) \to \frac{\lambda'}{\pi(1-q)}
\end{array}
\label{eq:13b}
\ee
Note the essential difference with \eq{eq:anderson}. It is easy to see that all tails of
intermediate monotonic sequences have scaling behavior as in the sequence $\rho_1^{S_2}(\lambda)$.

Return now to \eq{eq:23} and substitute for $k$ and $k'$ the asymptotic expansions
$k(\lambda)\approx \frac{\pi}{\sqrt{2-\lambda}}$ and $k'(\lambda')\approx \frac{\pi}{\lambda}$ (see
\eq{eq:tails2}). Since near the spectrum edges, $k\gg 1$ and $k\gg 1$, we get
\be
\sqrt{-\ln \left|\eta\left(\lambda + iy\right)\right|}\Big|_{{\lambda\to 2 \atop y\to 0}} \to
\frac{\sqrt{2-\lambda}}{\sqrt{12\pi y}} \quad \mbox{and} \quad \sqrt{-\ln \left|\eta\left(\lambda +
iy\right)\right|}\Big|_{{\lambda\to 0 \atop y\to 0}} \to \frac{\lambda}{\sqrt{12\pi y}}
\label{eq:24}
\ee
The tails in \eq{eq:24} exactly match the tails of the spectral density $\rho^A(\lambda)$ and
$\rho^B(\lambda)$ in \eq{eq:13b} if we make an identification $1-q = \sqrt{12\pi y}$.

The same analysis can be carried out for any monotonic sequence of peaks of the spectral density
depicted in \fig{fig:03}. This brings us to the central conjecture od this work, which uncovers
the number theoretic nature of the spectral density of Poissonian ensemble of large random
bi-diagonal symmetric matrices.

\noindent {\bf Conjecture.} Let $\rho_{\rm lin}(\lambda,q)$ be the spectral density of ensemble of
$N\times N$ bi-diagonal symmetric random Bernoulli matrices with the Poissonian distribution of
nonzero matrix elements defined in Eqs.\eq{eq:06}--\eq{eq:06a}. For $N\to\infty$ we have the
relation:
\be
\lim_{q\to 1^-} \frac{\rho_{\rm lin}(\lambda,q)}{\sqrt{-\ln
\left|\eta\left(\frac{1}{\pi}\arccos(-\lambda/2)+i\frac{(1-q)^2}{12\pi}\right)\right|}} = 1
\label{eq:25}
\ee
where $\eta$ is the Dedekind $\eta$-function defined in \eq{eq:14}. One can say, that Eq.\eq{eq:25}
establishes the link between spectral density of random one-dimensional Schr\"odinger-type
operators with the limiting values of modular function at rational points on the real axis.

\begin{figure}[ht]
\centerline{\includegraphics[width=15cm]{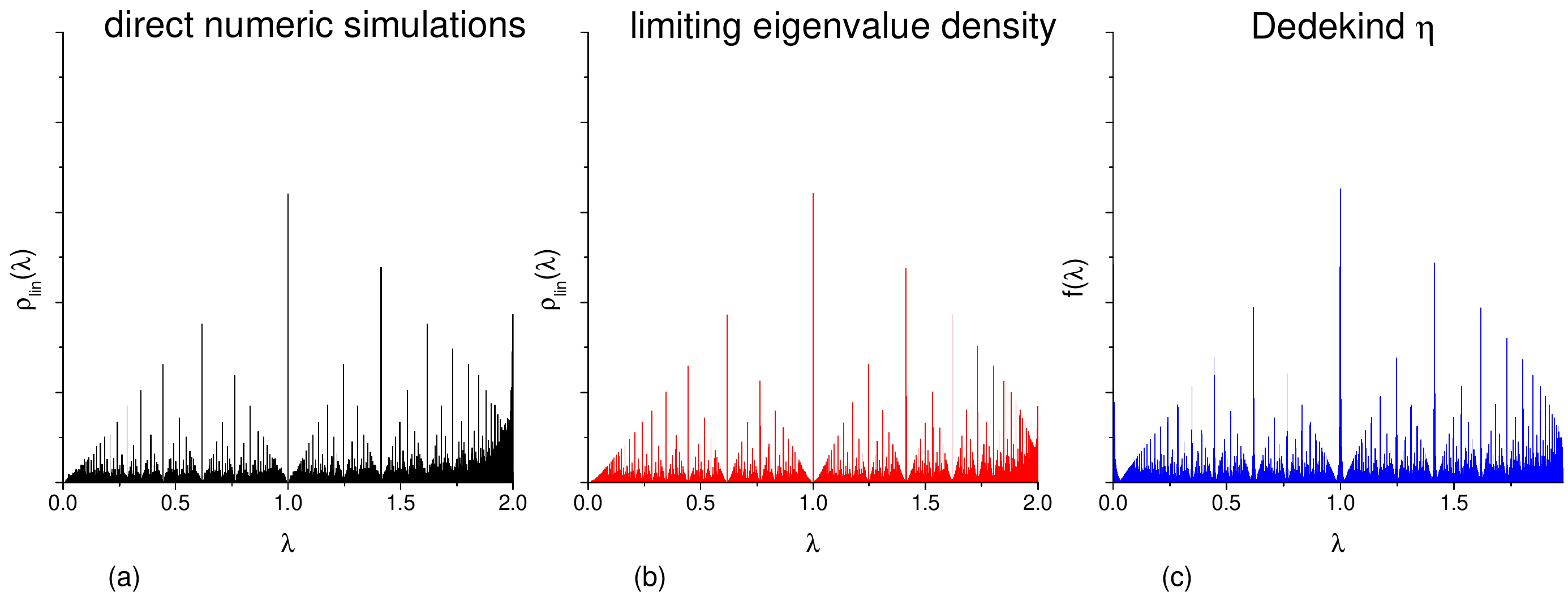}}
\caption{Spectral density $\rho_{\rm lin}(\lambda)$ of ensemble of bi-diagonal Poissonian random
matrices (shown only for $\lambda\ge 0$): a) direct numeric simulation; b) numeric visualization of
Eq.~\eq{eq:10}; c) analytic construction provided by Eq.~\eq{eq:25}.}
\label{fig:06}
\end{figure}

Empirical evidence in favor of the above conjecture is presented in \fig{fig:06} where we plot
three functions:

\begin{itemize}
\item[\fig{fig:06}a:] Numerically generated spectral density of ensemble of $N\times N$ ($N=3000$)
bi-diagonal Bernoulli random matrices $B$ (see \eq{eq:06}) with the Poissonian distribution of ``1"
at $q=0.98$;

\item[\fig{fig:06}b:] Numerically computed eigenvalue density $\rho(\lambda,q)$ given by Eq.\eq{eq:10}
for matrices of size $N=3000$ and $q=0.98$;

\item[\fig{fig:06}c:] Analytic expression given by the right-hand side of \eq{eq:25} at
$y=\frac{(1-q)^2}{12\pi}\approx 10^{-5}$.
\end{itemize}

\section{Discussion}

\subsection{Ultrametricity and rare-event statistics}

We investigated the spectral density of a sub-ensemble of sparse symmetric Bernoulli matrices near
the percolation point. We showed that in the vicinity of the percolation point linear chains
constitute the overwhelming majority of clusters---about 95\%. This led us to study the spectral
density of ensemble of linear chains only. The advantage of this consideration is the possibility
of exact treatment which allowed us to uncover the number-theoretic structure of the corresponding
density of states.

In a more practical setting, our analysis is applicable to investigation of polydisperse solutions
of linear polymers at any concentrations with Poissonian distribution in chain lengths, where the
ultrametricity could be directly seen in the absorption spectra. One tacit message coming from our
work is that experimenting with biological activity of highly diluted solutions of biologically
active substances, one should pay attention to a very peculiar structure of background noise
originated from the rare-event statistics of dissolved clusters. In order to make conclusion about
any biological activity of regarded chemical substance, the signal from background noise should be
clearly identified.

As concerns exact results, we showed that the spectral density, $\rho_{\rm lin}(\lambda)$, in the
ensemble of bi-diagonal random symmetric matrices with the Poissonian distribution of matrix
elements demonstrates a very peculiar ultrametric structure.

The derivation of the shape of the enveloping curve for the spectral density, $\rho_{\rm
lin}(\lambda)$, of random Shr\"odinger-type operators presented by random symmetric bi-diagonal
matrices, is based on simultaneous localization of eigenvalues, $\lambda_k=2\cos\frac{\pi}{k}$
($k=1,2,...$), constituting the principal sequence, and of their multiplicaticity (degeneracy),
$\rho(\lambda_k)=\frac{q^k}{1-q^k}$ ($0<q<1$). Eliminating $k$ from the parametric dependence
$(\lambda_k,\rho(\lambda_k))$, one gets the signature of a Lifshitz tail for Anderson localization
at the edge of the spectrum, $\rho(\lambda\to 2) \sim q^{-\pi/\sqrt{2-\lambda}}$, typical for the
one-dimensional Schr\"odinger random operators. We emphasize that in our case the derivation is
purely geometric.

Analyzing the limit  $q\to 1$, we have demonstrated that the function $\rho_{\rm lin}(\lambda)$
shares number-theoretic properties connected with the structure of the Dedekind $\eta-$function
near the real axis at rational points. Intriguingly, the limiting behavior of the Dedekind function
near real axis at rational points has tantalizing connections with Vassiliev knot invariants
\cite{zagier} and with quantum invariants of 3-manifolds \cite{zagier3}; a new concept of
\emph{quantum modular forms} was recently suggested \cite{zagier:quantum} as a potential
explanation.

\subsection{Manifestation of the Dedekind $\eta$-function in natural science}

Finally, we mention two other physical systems exhibiting a connection with the
$\eta$ function. In Ref.~\cite{voit} it has been argued that buckling of
a leaf of some plants (like lettuce or spinach) is related with the isometric embedding of hyperbolic graphs (Cayley
trees) into the 3D Euclidean space. This buckling is described by a function, $g(z)$, (Jacobian of
transformation), which defines the deformation of the hyperbolic graph in its projection from the
3D Euclidean space into the complex plane (see also the Section 5.1 of \cite{pomeau}).

\begin{figure}[ht]
\centerline{\includegraphics[width=15cm]{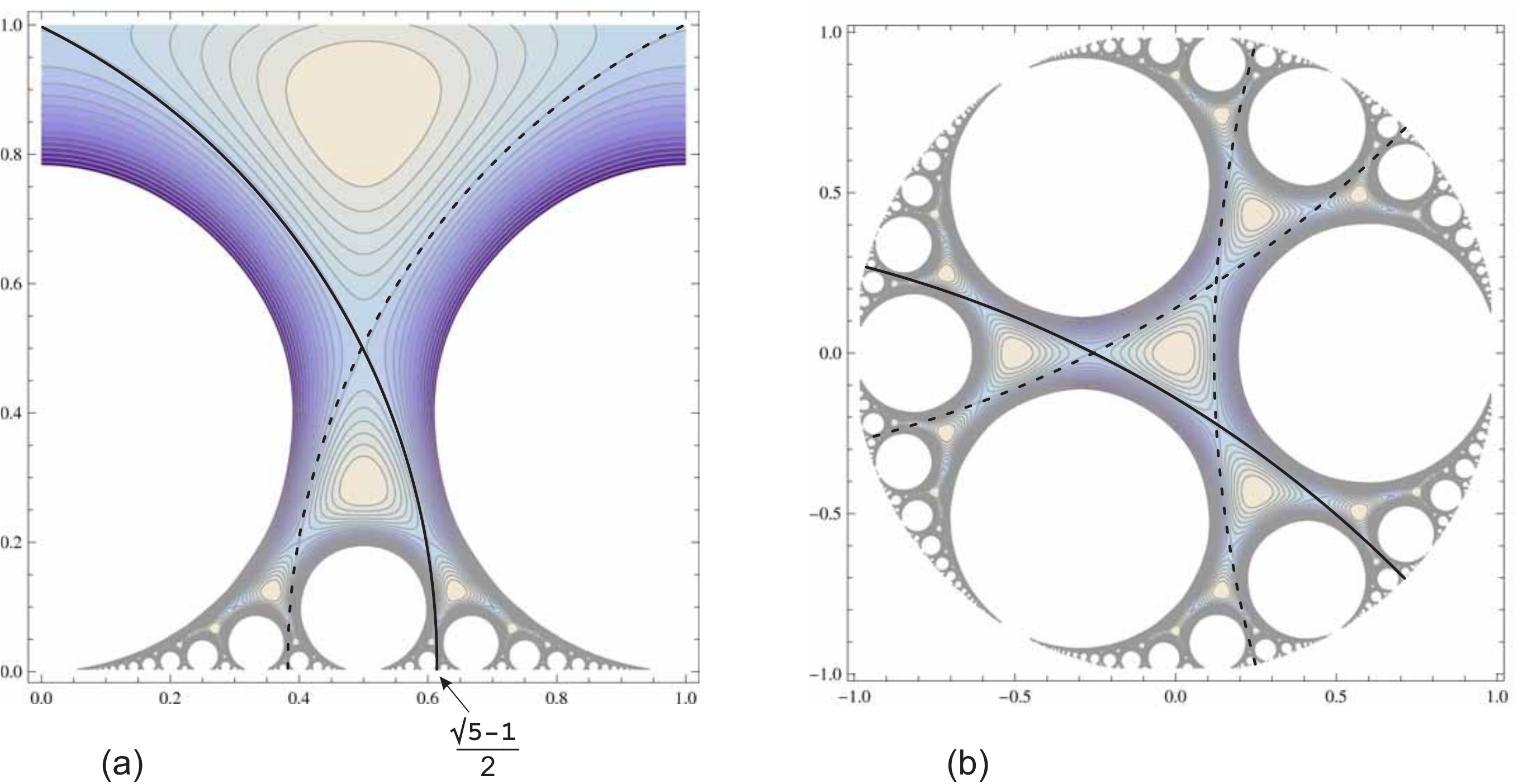}}
\caption{Cut ($f(z)\in[0.985, 1.0]$) of the contour plot of the function $f(z)$ (see
\eq{eq:15}): a) the representation in the upper half-plane, $z=x+iy$; b) the representation in the
unit disc obtained by the conformal mapping of the upper half-plant to the unit disc.}
\label{fig:07}
\end{figure}

For a regular 3-branching Cayley tree, the function $g(z)$ can be expressed via the Dedekind
$\eta$-function \cite{voit,vas}. It has been noted in \cite{vas} that properly normalized function
$g(z)$, namely, $f(z)=A^{-1} |\eta(z)| (\im z)^{1/4}$, defined in Eq.\eq{eq:15}, can be regarded as
a continuous tree isometrically embedded into the half-plane $\im z>0$. The constant $A$ in
\eq{eq:15} is chosen to set the maximal value of the function $f(z)$ to 1: $0<f(z)\le 1$ for any
$z$ in the upper half-plane $\im z > 0$. The function $f(z)$, being the modular function, has the
following property \cite{vas}: $f(z)$ has local maxima equal to 1 at the vertices of the
3-branching Cayley tree (and only of them) isometrically embedded into the upper half-plane $\im
z>0$. The cut of the function $f(z)\in [0.985,1.0]$ is shown in \fig{fig:07}a, while in
\fig{fig:07}b the same function is replotted in the unit disc, obtained via the conformal mapping
of the upper half-plane $\im z>0$ to the unit disc. Note the striking similarity of \fig{fig:07}a with the set of touching Ford circles (\fig{fig:farey}). In fact, the $x$-coordinates of centers of all lacunas in \fig{fig:07}a obey the
Farey construction.

Another well-known manifestation of number theory in natural science arises in the context of phyllotaxis
\cite{phyllo}. The connection of the cell packing with the Farey sequences has been observed long
time ago. It was not clear, however, why nature
selects the Fibonacci sequence among other possible Farey sequences. A tantalizing answer to
this question has been given by L. Levitov, see \cite{levitov}. He proposed an ``energetic" approach
to the phyllotaxis, suggesting that the development of a plant is connected with an effective
motion along the \emph{geodesics} on the surface associated with the energetic relief of growing
plant.

Intriguingly, the energy relief discussed in \cite{levitov} coincides with the relief of the
function $f(z)$ depicted in \fig{fig:07}. The black arcs drawn in \fig{fig:07}a,b are parts of
semicircles orthogonal to the boundaries and represent the geodesics of the surface $f(z)$, which
are \emph{open level lines} (by the symmetry, there are only two such level lines in \fig{fig:07}a
and three in \fig{fig:07}b). The best rational approximation of the boundary point,
$\frac{\sqrt{5}-1}{2}$, at which the solid geodesics in \fig{fig:07}a terminates, is given by the
continued fraction expansion
\be
\frac{\sqrt{5}-1}{2}=\frac{1}{\disp 1+\frac{1}{\disp 1+\frac{1}{\disp 1+...}}},
\label{eq:cont-fr}
\ee
cut at some level $k$. Cutting \eq{eq:cont-fr} at sequential $k$, we get the set of fractions
constituting the interlacing Fibonacci sequences, $\{F_k\}$, in nominators and denominators:
$$
\frac{1}{1},\; \frac{1}{2},\; \frac{2}{3},\; \frac{3}{5},\;
\frac{5}{8},...,\frac{F_{k-1}}{F_{k}},\frac{F_k}{F_{k+1}}
$$
Thus, the  normalized $\eta$-function, $f(z)$, plays a role of the energy relief of a growing
plant.

One can say that the Fibonacci numbers are the benchmarks, which fix the connection between
ultrametric and hyperbolic geometry, since, on one hand they appear as discrete symmetries of the
hyperbolic space, and on the other hand, being a subset of $p$-adic numbers, they uniquely
parameterize the distances in the ultrametric space \cite{avet-nech}.

\begin{acknowledgments}

We are grateful to Y. I. Manin and Y. Fyodorov for valuable discussions. SN acknowledges the
support of the IRSES DIONICOS grant, VA was supported by the Higher School of Economics program for
Basic Research, PLK thanks the Galileo Galilei Institute for Theoretical Physics and the INFN for
support during the completion of this work.

\end{acknowledgments}

\end{document}